\documentclass{article}

\usepackage{fullpage,comment, amsmath, amsthm, amssymb, cleveref, url}
\usepackage[T1]{fontenc}
\usepackage[utf8]{inputenc}
\usepackage{lmodern}
\usepackage{footnote}
\usepackage{xcolor}
\usepackage{thmtools}
\usepackage{thm-restate}
\usepackage{cleveref}
\usepackage{xpatch}

\makeatletter
\xpatchcmd{\thmt@restatable}% Edit \thmt@restatable
   {\csname #2\@xa\endcsname\ifx\@nx#1\@nx\else[{#1}]\fi}% Replace this code,
   {\ifthmt@thisistheone% ... with this, which only does so at the original call site.
       \csname #2\@xa\endcsname\ifx\@nx#1\@nx\else[{#1}]\fi%
       \else\fi}
   {}{\typeout{FIRST PATCH TO THM RESTATE FAILED}} % execute on success/failure 
\xpatchcmd{\thmt@restatable}% A second edit to \thmt@restatable, so that we also
   {\csname end#2\endcsname}
   {\ifthmt@thisistheone\csname end#2\endcsname\else\fi}%
   {}{\typeout{FAILED SECOND THMT RESTATE PATCH}}
\makeatother

\newcommand{\ppa}{\mathbb{P}_A}
\newcommand{\ppg}{\mathbb{P}_g}
\newcommand{\ppgt}{\mathbb{P}_{\tilde{g}}}
\newcommand{\ppp}{\mathbb{P}}
\newcommand{\eps}{\varepsilon}
\newcommand{\E}{\mathbb{E}}

\newcommand{\Normal}{N}
\newcommand{\norm}[1]{\left\lVert#1\right\rVert^2}
\newcommand{\normos}[1]{\left\lVert#1\right\rVert}
\renewcommand{\log}{\lg}

\newtheorem{theorem}{Theorem}
\newtheorem{lemma}[theorem]{Lemma}

\newtheorem{definition}[theorem]{Definition}

\newcommand{\Reals}{\mathbb{R}}
\newcommand{\alert}[1]{\textbf{\color{green}
[#1]}\marginpar{\textbf{\color{green}**}}\typeout{ALERT:
\the\inputlineno: #1}}

\renewcommand{\Pr}{\ppp}
\newcommand{\FastJL}{ScaledJL($\eps,\delta,s$)-matrix}

\title{Barriers for Faster Dimensionality Reduction}
\author{Ora Nova Fandina \thanks{{\tt fandina@cs.au.dk}. Aarhus
    University. Supported by Independent Research Fund Denmark (DFF) Sapere Aude Research Leader grant
No 9064-00068B.}
 \and Mikael M\o ller H\o gsgaard\thanks{{\tt hogsgaard@cs.au.dk}. Aarhus
    University. Supported by Independent Research Fund Denmark (DFF) Sapere Aude Research Leader grant
No 9064-00068B.}
 \and Kasper Green Larsen\thanks{{\tt larsen@cs.au.dk}. Aarhus
    University. Supported by Independent Research Fund Denmark (DFF) Sapere Aude Research Leader grant
No 9064-00068B.}
}

\begin{document}
 
\date{}
\maketitle

\begin{abstract}
The Johnson-Lindenstrauss transform allows one to embed a dataset of $n$ points in $\mathbb{R}^d$ into $\mathbb{R}^m,$ while preserving the pairwise distance between any pair of points up to a factor $(1 \pm \varepsilon)$, provided that $m = \Omega(\eps^{-2} \lg n)$. The transform has found an overwhelming number of algorithmic applications, allowing to speed up algorithms and reducing memory consumption at the price of a small loss in accuracy. A central line of research on such transforms, focus on developing fast embedding algorithms, with the classic example being the Fast JL transform by Ailon and Chazelle. All known such algorithms have an embedding time of $\Omega(d \lg d)$, but no lower bounds rule out a clean $O(d)$ embedding time. In this work, we establish the first non-trivial lower bounds (of magnitude $\Omega(m \lg m)$) for a large class of embedding algorithms, including in particular most known upper bounds.
\end{abstract} 

\thispagestyle{empty}
\newpage
\setcounter{page}{1}
\section{Introduction}
Working with high dimensional data can be both costly in memory and computational power, motivating the study of dimensionality reduction techniques. The goal of dimensionality reduction is to take a high dimensional dataset $X$ and embed it to a dataset $Y$ in a lower dimensional space. If $Y$ approximately preserves similarities between points in $X$, then one may use $Y$ as input to an algorithm in place of $X$ to save both memory and computation time at the cost of a small inaccuracy in ones output.
A greatly celebrated dimensionality reduction result is the Johnson-Lindenstrauss lemma \cite{JLlemma}, which states: For any fixed $X\subset\mathbb{R}^d$, with the size of $X$ being $n$, and any distortion $0<\eps<1$, there exists a map $f: X\rightarrow \mathbb{R}^m$ such that for all $x,y\in X$
$$
\|f(x)-f(y)\|_2\in (1\pm\eps)\|x-y\|_2,
$$
with $m$ being $\Theta(\eps^{-2}\log n)$~\cite{JLlemma,DBLP:conf/focs/LarsenN17}.
Thus the mapping is approximately preserving the Euclidean distances between the points in $X$ in the lower dimensional space $\mathbb{R}^m$. The property of preserving pairwise distances via the Johnson-Lindenstrauss lemma have found great use in many applications, for instance as a preprocessing step to speed up machine learning algorithms. 

A standard approach for obtaining an embedding $f$ satisfying the above, is to pick a random $m \times d$ matrix $A$ with each entry being i.i.d. $\Normal(0,1)$ distributed~\cite{IM98} (or uniform $-1/1$~\cite{rademachers}) and embedding any input $x \in X$ to $f(x)= m^{-1/2}Ax$. Computing such an embedding thus takes $O(md)$ time. In some applications of dimensionality reduction, this becomes the bottleneck in the running time, thus motivating faster embedding algorithms. The work on faster dimensionality reduction in Euclidian space can be divided roughly into two categories: 1) using sparse embedding matrices $A$, or 2), using matrices $A$ with special structure that allows fast matrix-vector multiplication. In both cases, the fastest embedding algorithms use super-linear time in the input dimensionality in the worst case. For sparse matrices, there is near-tight lower bound by Nelson and Nguyen ~\cite{DBLP:conf/stoc/NelsonN13} showing that the embedding time cannot be reduced below roughly $\Omega(d \eps^{-1} \lg n)$. For structured matrices, the fastest embeddings use at least $\Omega(d \lg m)$ time, however in this case there are no lower bounds ruling out faster embeddings that could conceivably embed a vector in $O(d)$ time see e.g. \cite{DBLP:journals/corr/abs-2204-01800,DBLP:journals/corr/abs-2003-10069}. Working towards such lower bounds is the focus of this work.

\paragraph{Our Contributions.}
In this work, we establish the first non-trivial lower bounds on the time required for dimensionality reduction in Euclidian space when not restricted to using sparse matrices to perform the embedding. Focusing on the case of $d= c m$, for a constant $c>1$ and optimal $m = O(\eps^{-2} \lg n)$, we prove that a \emph{large class} of embedding algorithms, including most known upper bounds, must use time $\Omega(m \lg m)$. This coincides with known upper bounds for several tradeoffs between $\eps$ and $n$. In addition to establishing a first lower bound, we believe our careful definition of the class of algorithms that the lower bound applies to, shines light on the barriers faced when developing fast embedding algorithms.

In the following section, we survey previous work and formally present our results.

\subsection{Fast Dimensionality Reduction}
As mentioned above, the previous work on fast dimensionality reduction can be divided into two categories, either based on sparse matrices or on structured matrices. We elaborate on these approaches in the following.

\paragraph{Sparse JL.} The basic idea in sparse JL embeddings, is to use an embedding matrix $A$ with only $s < m$ non-zeros per column. With such a matrix $A$, the product $Ax$ can be computed trivially in $O(sd)$ time rather than $O(md)$, thus speeding up the embedding. Moreover, if $x$ itself has few non-zeros, then the product may even be computed in $O(s \|x\|_0)$ time, where $\|x\|_0$ is the number of non-zeros in $x$. Using sparse embedding matrices was initiated by~\cite{DBLP:journals/jcss/Achlioptas03} and culminated with the current state-of-the-art embedding by Kane and Nelson~\cite{DBLP:journals/jacm/KaneN14} who showed that it suffices to pick a matrix $A$ having $s = O(\eps^{-1} \lg n)$ random entries (without replacement) in each column set uniformly and independently to $-1/1$ and embedding a vector $x$ to $s^{-1/2}Ax$. Moreover, this nearly matches a sparsity lower bound by Nelson and Nguyen~\cite{DBLP:conf/stoc/NelsonN13} who showed that any sparse embedding matrix must have $s=\Omega(\eps^{-1}\lg n/\lg(1/\eps))$ non-zeros per column. Another line of research in this direction, studies sparsities $s$ below the lower bound by Nelson and Nguyen. For instance, Feature Hashing~\cite{WeinbergerDLSA09} considers the extreme case of $s=1$. Of course, such embeddings cannot work for all data sets $X$. However, as shown by Weinberger et al.~\cite{WeinbergerDLSA09} and later refined by Kamma et al.~\cite{FeatureHashing} and generalized to $s>1$ by Jagadeesan~\cite{DBLP:conf/nips/Jagadeesan19}, one can use extremely sparse embedding matrices, provided that for all pairwise difference vectors $z = x-y$ for $x,y \in X$, the ratio $\|z\|_\infty/\|z\|_2$ is small. That is, there are no single large coordinates in $z$.

\paragraph{Fast JL.}
The second line of research on fast embeddings exploits structured matrices $A$ with fast matrix-vector multiplication algorithms. Ailon and Chazelle~\cite{Ailon2009TheFJ} initiated this direction by introducing the FastJL transform. FastJL embeds a vector by computing a product $m^{-1/2}PHDx$, where $P$ is a sparse matrix, $H$ is the normalized $d \times d$ Hadamard matrix and $D$ is a diagonal matrix with random signs on the diagonal. The trick is that computing $Dx$ can be done in $O(d)$ time and computing $H(Dx)$ takes only $O(d \lg d)$ time by exploiting the structure of the Hadamard matrix. Finally, the transformation $HDx$ has the effect of ``smoothening'' out the coordinates of the input vector, making the ratio $\|HDx\|_\infty/\|HDx\|_2$ small. This is precisely the setup allowing very sparse embedding matrices. Concretely, Ailon and Chazelle~\cite{Ailon2009TheFJ} showed that it suffices to let each entry in $P$ be non-zero with probability $q = O((\lg^2n)/d)$, resulting in a total embedding time of $O(d \lg d + m \lg^2 n)$. Their analysis was recently refined by Fandina et al.~\cite{DBLP:journals/corr/abs-2204-01800}, showing that the sparsity parameter $q$ can be reduced further. Numerous other embeddings exploiting structured matrices has since then been introduced~\cite{DBLP:journals/corr/abs-1009-07441, 4959960, dualBCH,Bamberger2017OptimalFJ}, including for instance embeddings based on Toeplitz matrices~\cite{https://doi.org/10.1002/rsa.20360,article1,DBLP:journals/algorithmica/FreksenL20} and the Kac random walk~\cite{kac,DBLP:journals/corr/abs-2003-10069}. If one insists on optimal $m = O(\eps^{-2} \lg n)$ dimensions in the embedding, then the current state-of-the-art is either the FastJL transform or the Kac random walk depending on the relationship between $n$ and $\eps$. However none of these are faster than $O(d \lg m)$ for any tradeoff between $\eps$ and $n$.

Unlike the sparse matrix case, there are no known lower bounds ruling out e.g. $O(d)$ time embeddings via structured matrices. Naturally, the reason for this, is that it is much harder to prove lower bounds for general embedding algorithms that exploit structured matrices than merely bounding the sparsity of the embedding matrix. In fact, proving super-linear lower bounds for general linear circuits (which capture current embedding algorithms) is a major open question in complexity theory. In light of this obstacle, which we will elaborate on in Section~\ref{sec:barriers}, we identify common traits in most known upper bounds that we exploit to prove lower bounds for dimensionality reduction. In the following, we formally define the model under which we prove our lower bound.

\subsection{Formal Lower Bound}
As mentioned earlier, our lower bound holds for a large class of dimensionality reducing maps. This class is captured by a certain scaling parameter. Concretely, we define a ScaledJL-matrix as follows:
\begin{definition}
  \label{def:fastjl}
Let $0<\eps,\delta<1$ and $s\in\mathbb{N}$. A stochastic matrix $A\in\mathbb{R}^{m\times d}$ is said to be a \FastJL, if for any $x\in\mathbb{R}^d$ we have that
\begin{gather*}
\ppa\left[\norm{s^{-1/2}Ax}_2 \not\in(1\pm\varepsilon)\norm{x}_2\right]<\delta.
\end{gather*}
\end{definition}
Let us remark a few things about Definition~\ref{def:fastjl}. First, we assume that a \FastJL\ $s^{-1/2}A$ preserves the (squared) norm of any single vector $x$ up to $(1\pm \eps)$ except with probability $\delta$. This is the standard definition of a distributional Johnson-Lindenstrauss transform and all known upper bounds give such a guarantee. In greater detail, known upper bounds prove the distributional guarantee and then sets $\delta < 1/n^2$ and applies a union bound over all $z = x-y$ for $x,y \in X$ to conclude that the embedding preserves all pairwise (squared) distances among vectors in $X$. In this work, we focus on the squared distance as it simplifies calculations and anyways only changes $\eps$ by a constant factor. The non-standard thing in Definition~\ref{def:fastjl} is the scaling parameter $s$. Of course, such a scaling parameter can also be implicitly hidden in $A$ by scaling all entries of $A$ by $s^{-1/2}$. To explain the role of $s$ in our model, we need to first introduce a linear circuit/algorithm as defined e.g. by Morgenstern:
\begin{definition}\cite{MORGENSTERN197159}
  \label{def:linear}
A linear algorithm takes as an input $1,x_1,\ldots,x_d \in \mathbb{R}$ and proceeds in $t>0$ steps. In the $l$'th step the algorithm computes $x_{d+l}$ by $x_{d+l}=\lambda_{d+l} x_{j}+\mu_{d+l} x_{i}$ for some pair of indices $i,j<d+l$, where $\lambda_{d+l},\mu_{d+l}\in  \mathbb{R}$. 

We say that a linear algorithm computes a linear transformation $B\in\mathbb{R}^{m\times d}$ if there exist indices $1\leq k_1,\ldots,k_m\leq d+t$ such that: $(Bx)_1=x_{k_1},\ldots,(Bx)_m=x_{k_m}$ for every possible input $x=(x_1,\ldots,x_d)\in\mathbb{R}^d$.
\end{definition}

Note that the number of steps $t$ determines the number of operations performed by the algorithm (up to a factor 3).
Proving super-linear lower bounds for linear algorithms in the sense of Definition~\ref{def:linear}, is a major open problem~\cite{valiantRigidity}. Thus several previous works~\cite{Morg73,doi:10.1137/S0097539794275665} have considered restrictions where the coefficients $\lambda$ and $\mu$ are bounded in absolute value by a constant $r$ independent of $m$ and $d$. This is crucially necessary if one wants to avoid the long-standing complexity theoretic barriers further elaborated on in Section~\ref{sec:barriers}.

With this in mind, the role of $s$ in our definition of \FastJL\ becomes clearer. Concretely, if we consider an embedding $s^{-1/2}Ax$, then we think of $A$ as being computable by a linear algorithm/circuit where all coefficients $\lambda_i$ and $\mu_i$ are bounded by a constant. This naturally leads to a scaling factor $s^{-1/2}$ for some $s$. Such a scaling also occurs in most known upper bounds. Let us first state our main lower bound result and then discuss how it relates to known constructions:
\begin{theorem}\label{circuitJLsize}
Let $A\in\mathbb{R}^{m\times d}$ be a \FastJL\ for $\varepsilon\leq1/4$, $\delta\leq C$ ($C$ being some universal constant), $s\in\mathbb{N}$, $m = \Theta(\varepsilon^{-2}\log(1/\delta))$ and $d \geq m$, then the expected (over the random choice of $A$) minimum number of operations needed for any linear algorithm computing $A$ with $|\lambda_i|,|\mu_i| \leq 1$ for all $i$ is $\Omega(m\log s)$.
\end{theorem} 
Let us briefly argue that most known constructions are of the form captured by the lower bound and the definition of a \FastJL. Concretely, these upper bounds have $\log s = \Omega(\log m)$ and thus our lower bound shows that it must take $\Omega(m \lg m)$ operations to compute these embeddings, even if more clever linear algorithms could be devised. As an example of an upper bound, consider first the classic JL construction using a matrix $A$ with i.i.d. random $-1/1$ entries and a scaling of $s^{-1/2}=m^{-1/2}$. In this case, the matrix $A$ can clearly be computed by a linear algorithm using coefficients bounded by $1$ in absolute value (just carry out the trivial algorithm). So it falls under the definition of a \FastJL\ with $s = m$. Next consider embeddings based on Toeplitz matrices~\cite{https://doi.org/10.1002/rsa.20360,article1,DBLP:journals/algorithmica/FreksenL20}. Here we embed as $m^{-1/2}TDx$, where $D$ is a diagonal with random signs and $T$ is a Toeplitz matrix with random signs on its diagonals. The matrix $T$ can be computed via a fast Fourier transform using coefficients bounded by a constant. Hence the construction also falls under the definition of \FastJL\ with $s = m$. We could also consider the sparse JL transform by Kane and Nelson~\cite{DBLP:journals/jacm/KaneN14}. Their construction uses an embedding matrix where each column has $t = \Theta(\eps^{-1} \lg n)$ non-zero entries, each of magnitude $t^{-1/2}$. Such a sparse embedding is typically computed by moving the scaling $t^{-1/2}$ outside and then doing the straight-forward sparse matrix-vector multiplication using constant magnitude coefficients. It thus falls under the definition of a \FastJL\ with $s = t = \Theta(\eps m)$. This has $\lg s = \Omega(\lg m)$ when $m$ is optimal $O(\eps^{-2} \lg n)$.
Finally, consider for instance the $m^{-1/2}PHD$ construction by Ailon and Chazelle~\cite{Ailon2009TheFJ}. They use the \emph{normalized} Hadamard matrix, i.e. all entries in $H$ are scaled down by $d^{-1/2}$. If we move that scaling factor outside, as $(md)^{-1/2}P\bar{H}D$, then $\bar{H}$ is computed recursively using coefficients of $1$ and $-1$. The entries of $P$ are $b \cdot \Normal(0,q^{-1})$ distributed, where $b$ is a Bernoulli random variable with success probability $q$ for a $q > \lg(1/\delta)/d$. With high probability, no entry of $P$ is thus larger than about $O(\sqrt{d})$. Moving this scaling factor outside, it cancels out with the $d^{-1/2}$ from the Hadamard matrix and then $P$ can also be computed using coefficients bounded by a constant and the final algorithm is a \FastJL\ with $s = \Theta(m)$. Common to all these approaches, is that they project onto something that resembles a random $m$-dimensional subspace. Intuitively, such a matrix should have $m$ rows all of norm about $\sqrt{d/m}$. With $d$ columns, this would imply that each entry should be about $m^{-1/2}$ in magnitude. Moving the scaling factor outside to have constant magnitude entries, results in the $m^{-1/2}$ scaling factor observed in all these upper bounds.
 
Thus many known upper bounds fall under the definition of a \FastJL\ with a scaling $s$ satisfying $\lg s = \Omega(\lg m)$. Theorem~\ref{circuitJLsize} therefore sheds light on why they all require $\Omega(m \lg m)$ time (which is $\omega(d)$ when $d = O(m)$). Let us also mention the only upper bound we are aware of, that does not seem to suffer from the lower bound. In the Kac JL transform~\cite{kac,DBLP:journals/corr/abs-2003-10069}, one embeds a vector by repeatedly picking two random coordinates, among the $d$ input coordinates, and performing a random rotation on the two. After sufficiently many steps ($\Omega(d \lg d + m \lg n)$ in the current analysis), all but the first $m$ coordinates are discarded and those $m$ coordinates are scaled by $\sqrt{d/m}$. While seemingly not being captured by the lower bound, we remark that the analysis of Kac JL cannot be sharpened to $o(d \lg d)$ steps as otherwise, by a coupon collector argument, there is a vector $e_i$ among $e_{m+1},\dots,e_d$ whose coordinate $i$ is never involved in a rotation and hence $e_i$ is embedded to $0$.

Of course, it would have been more natural, if our lower bound in Theorem~\ref{circuitJLsize} only required bounded coefficients in the linear algorithm, not that there is also a scaling parameter $s^{-1/2}$. Unfortunately, as we argue in Section~\ref{sec:barriers}, it seems unlikely that we can establish such a lower bound using current techniques. We thus believe our results can be seen in two ways: 1), as providing strong evidence that FastJL constructions cannot be made much faster, or 2), as pointing towards a direction for further improvements, by trying to design embeddings where a constant scaling parameter $s$ suffices, or super-constant coefficients are used when computing the embedding, or perhaps using non-linearity. 

\subsection{Barriers for Linear Algorithm Lower Bounds}
\label{sec:barriers}
Proving super-linear unconditional lower bounds is one of the biggest barriers in many areas of complexity theory, including in particular for linear operators. A natural computational model for computing linear operators is a linear algorithm, a.k.a. linear circuit, as in Definition~\ref{def:linear}. While being a very natural model of computation for linear operators, capturing in particular all known JL constructions, it suffers from a lack of tools for proving lower bounds (without any assumptions on coefficients). Concretely, there are still no super-linear size lower bounds, even for circuits restricted to logarithmic depth. Moreover, this road block is not for lack of trying. For instance, already in 1977, Valiant~\cite{valiantRigidity} introduced the notion of \emph{matrix rigidity}. Loosely stated, the rigidity of a square matrix (corresponding to a linear operator) $A \in \Reals^{n \times n}$, is the minimum number of entries in $A$ that needs to be changed to reduce its rank below $n/2$. Valiant showed that any explicit matrix $A$ with rigidity $\Omega(n^2/\lg \lg n)$ cannot have a linear-sized and log-depth linear circuit for computing the corresponding linear operator. Matrix rigidity has since then been the topic of much research, see e.g.~\cite{noteRigid,DBLP:conf/approx/AlonPY09,DBLP:conf/coco/SarafY11,omriRigid}, however none of these works lead to super-linear lower bounds (also when considering rectangular matrices) for explicit matrices, despite the fact that a random matrix has high rigidity with high probability.

\paragraph{Bounded Coefficients.}
In light of the above strong barriers for proving lower bounds for linear circuits, a natural restriction to the computational model, is to assume that all coefficients $\lambda_i$ and $\mu_i$ used by the gates are bounded in absolute value by a constant $r$. Indeed, if we enforce such a restriction, then Morgenstern~\cite{Morg73} for instance proved an $\Omega(n \lg n)$ lower bound on the size of any linear circuit computing the $n \times n$ \emph{unnormalized} fast Fourier transform. Similarly, Chazelle~\cite{doi:10.1137/S0097539794275665} proved $\Omega(n \lg n)$ lower bounds for linear circuits, with bounded integer coefficients, for computing linear transformation corresponding to incidence matrices for various geometric range searching problems. Common to these techniques, is that they relate the circuit complexity to the eigenvalues of the corresponding matrix $A$. In particular, the lower bounds one obtains peak at $\Omega(\ell \lg \gamma_\ell)$, where $\gamma_\ell$ denotes the $\ell$'th largest eigenvalue of $A^TA$.

Now in the context of dimensionality reduction, an embedding matrix $A \in \Reals^{m \times d}$ can have at most $m$ non-zero eigenvalues. This means that lower bounds obtained via these techniques will be proportional to only $\Omega(m \lg \gamma_\ell)$ for an $\ell \in \Theta(m)$. Since the size of the circuit is already at least $d$, it makes most sense from a lower bound point of view to consider setups where $m$ and $d$ are within constant factors. However, since embedding matrices $A$ must preserve the norm of standard unit vectors $e_i$, their columns will have norms of magnitude $(1 \pm \eps)$. This implies that the trace of $A^TA$ is $d(1 \pm \eps) = \Theta(m)$. Since the trace of $A^TA$ equals the sum of its eigenvalues, we get for $\ell \in \Theta(m)$ that $\gamma_\ell$ is
at best a constant. Thus the lower bounds we may hope to obtain are only $\Omega(m)$, i.e. trivial. Thus considering only the restriction to have coefficients bounded by a constant is insufficient for proving non-trivial lower bounds using known techniques.

\paragraph{Output Scaling.}
Having observed the above, we examined existing FastJL constructions and found a common trait in most of them: they embed a vector $x$ by computing $s^{-1/2}Ax$ for some scaling factor $s$ and matrix $A$, where $A$ can be computed efficiently by a linear circuit using coefficients of constant magnitude. Given the obstacles mentioned above, we thus settled on proving lower bounds for embeddings that follow this template, resulting in Theorem~\ref{circuitJLsize} above. 
\section{Lower Bound for Linear Algorithms}
The goal of this section is to prove our lower bound from \Cref{circuitJLsize} on the operations needed for any linear algorithm computing a \FastJL. We state a stronger version of the theorem here:
\begin{theorem}\label{circuitJLsize1}
Let $A\in\mathbb{R}^{m\times d}$ be a \FastJL\ for $\varepsilon\leq1/4$, $\delta\leq C$($C$ being some universal constant), $s\in\mathbb{N}$ and $t \varepsilon^{-2}\log(1/\delta) = m$, $t\geq1$ and $d\geq m$, then the expected (over the random choice of $A$) minimum number of operations needed for any linear algorithm computing $Ax$ for any $x\in \mathbb{R}^d$ with $|\lambda_i|,|\mu_i|<r$ for all $i$ and $r>1/2$, is $\Omega(m\log(s/t^2)/(t\log(2r))$.
\end{theorem}
We notice that \Cref{circuitJLsize} is a special case of \Cref{circuitJLsize1} where $r$ is set equal to $1$ and $t = \Theta(1)$.

The main tool for proving \Cref{circuitJLsize1} is a lemma by Morgenstern relating the operations needed by a linear algorithm computing a linear transformation $B$, to the determinants of square submatrices of $B$:

\begin{lemma}\label{morgensternlemma} \cite{Morg73}
Let $B$ be a real matrix and let $\Delta(B)$  denote the maximum over the absolute value of the determinant of any square submatrix of $B$. A linear algorithm computing the linear transformation $B$, with $|\lambda_i|,|\mu_i|<r$ for all $i$ and $r>1/2$, must use at least $\log (\Delta(B))/\log (2 r)$ operations.
\end{lemma}

Using \Cref{morgensternlemma} as our offset, our goal is thus to show that any \FastJL\ $A$ must have a submatrix whose determinant is in the order of $s^{\Omega(m)}$. Since $A$ is allowed to be stochastic and fail to preserve the norm of a vector $x$ with probability $\delta$, we only prove that this holds with constant probability over $A$:

\begin{lemma}\label{masterlemma}
Let $A\in\mathbb{R}^{m\times d}$ be a \FastJL\ for $\varepsilon\leq1/4$, $\delta\leq C$ ($C$ being some universal constant), $s\in\mathbb{N}$ and $t \varepsilon^{-2}\log(1/\delta)= m$, $t\geq1$ and $d\geq m$, then there exist a set $S\subseteq \text{supp}(A)$ such that $\ppa\left[S\right]\geq 1/2$ and for $B\in S$ it holds that there exists a square submatrix $F$ of $B$ such that 
\begin{gather*}
|\det(F)|\geq \left(c^2s/(3(et)^2)\right)^{\lceil cm/t\rceil/2}
\end{gather*}
where $c$ is some universal constant less than $1$.
\end{lemma}

The proof of \Cref{circuitJLsize1} follows immediately from the above two lemmas:

\begin{proof}[Proof of \Cref{circuitJLsize1}] Let $A$ be a \FastJL. \Cref{masterlemma} gives the existence of a set $S \subseteq \text{supp}(A)$ with $\Pr_A[S] \geq 1/2$ and for $B\in S$, $B$ has a square submatrix $F$ such that $|\det(F)|\geq \left(c^2s/(3(et)^2)\right)^{\lceil cm/t\rceil/2}$ implying that $\Delta(B)\geq \left(c^2s/(3(et)^2)\right)^{\lceil cm/t\rceil/2}$. It now follows by \Cref{morgensternlemma} that a linear algorithm calculating $Bx$ for all $x\in\mathbb{R}^d$ must use $\log(\Delta(B))/\log(2r)$ operations. Since $\log(\Delta(B))\geq (\lceil cm/t\rceil)\log(c^2s/(3(et)^2))/2=\Omega(m\log(s/t^2)/t)$ we get that $\log(\Delta(B))/\log(2r)=\Omega(m\log(s/t^2)/(t\log(2r))$. Thus we conclude, since $\Pr_A[S]\geq 1/2$, that the expected number of operations needed by any linear algorithm computing the transformation $A$ is $\Omega(m\log(s/t^2)/(t\log(2r))$, which concludes the proof of \Cref{circuitJLsize1}.
    \end{proof}

The main challenge we face is thus establishing \Cref{masterlemma}, i.e. proving that for any \FastJL\ $A$, it is often the case that $A$ has a square submatrix of large determinant. This is the focus of the next section.

\subsection{Submatrix with Large Determinant (Proof \Cref{masterlemma})} \label{masterlemmaproofsection}
To prove \Cref{masterlemma}, we have to show that with probability at least $1/2$, a \FastJL\ has a square submatrix with an $\left(c^2s/(3(et)^2)\right)^{\lceil cm/t\rceil/2}$ large determinant. For this, we will use a technical lemma from~\cite{DBLP:conf/stacs/Larsen19} which relates the eigenvalues of $B^TB$ to the determinants of square submatrices of $B$:

\begin{lemma}\label{lemmaeigenvalues1} (\cite{DBLP:conf/stacs/Larsen19} proof of Theorem 10) For $B\in\mathbb{R}^{m\times d}$, with $m\leq d$, let $\lambda_{1} \geq \lambda_{2} \geq \cdots \geq \lambda_{m} \geq 0$ denote the eigenvalues of $B^{T} B$. For all positive integers $l \leq m$, there exists a square submatrix $F \in \mathbb{R}^{l \times l}$ of $B$ such that $$|\operatorname{det}\left(F\right)|\geq \sqrt{\frac{\prod_{i=1}^l\lambda_i}{\binom{d}{l} \binom{m}{l}}}.$$
\end{lemma}
By the above lemma, we can reduce the problem of finding a square submatrix of a \FastJL\ $A$ with large determinant, to lower bounding the eigenvalues of a $A^TA$. Using $\lambda_i(B^TB)$ to denote the $i$'th largest eigenvalue of $B^TB$, this is precisely the contents of the following lemma:

\begin{lemma}\label{lemma2}
Let $A\in\mathbb{R}^{m\times d}$ be a \FastJL\ for $\varepsilon\leq1/4$,  $\delta\leq C$($C$ being some universal constant) and $s\in\mathbb{N}$, $t \varepsilon^{-2}\log(1/\delta)= m$, $t\geq1$ and $d\geq m$, then there exist a set $S\subseteq \text{supp}(A)$ such that $\ppa\left[S\right]\geq 1/2$ and for $B\in S$ it holds that
\begin{align*}
\lambda_{\left\lceil c m/t \right \rceil}
    (B^TB)\geq ds/(3m)
\end{align*}
where $c$ is some universal constant less than $1$.
\end{lemma}
Before we give the proof of \Cref{lemma2}, let us see that it suffices to finish the proof of \Cref{masterlemma}:
\begin{proof}[Proof of \Cref{masterlemma}]
Let $A$ be \FastJL\ such that the conditions of \Cref{lemma2} are met. We then have for $B$ in the set $S$ described in \Cref{lemma2} that the $l=\lceil cm/t \rceil$'th largest eigenvalue of $B^TB$ is at least $ds/(3m)$. Now by \Cref{lemmaeigenvalues1}. we have that there exist a square submatrix $F\in\mathbb{R}^{l\times l}$ of $B$ such that $|\det(F)|\geq( \prod_{i=1}^l\lambda_i / \binom{d}{l} \binom{m}{l})^{1/2}$. Now using these two properties combined with $\binom{n}{k}\leq (en/k)^k$ and $ l\geq cm/t$ we get that
$$|\det(F)|\geq \left (\prod_{i=1}^l\lambda_i/\left (\binom{d}{l} \binom{m}{l}\right )\right )^{1/2}
\geq 
\left(dsl^2/(3e^2dm^2)\right)^{l/2}
\geq
\left(c^2s/(3(et)^2)\right)^{\lceil cm/t\rceil/2}.$$
Thus  \Cref{masterlemma} follows by the conditions in \Cref{masterlemma} and \Cref{lemma2} on the \FastJL\ being the same.
\end{proof}

After having established the above connection between eigenvalues and linear algorithms, we are left with proving \Cref{lemma2}, i.e. to show that for a \FastJL\ $A$, it is often the case that $A^TA$ has many large eigenvalues. We first give an overview of the main ideas in the proof, before proceeding to give the formal details.

\paragraph{Proof Overview.}
The proof of \Cref{lemma2} is at a high level inspired by methods used in \cite{mlowerboundGreenNelson}. The main result of \cite{mlowerboundGreenNelson} was a lower bound of $m = \Omega(\eps^{-2}\lg n)$ on the embedding dimension of any linear dimensionality reducing map. Their lower bound was proved for a ``hard'' set of vectors consisting of the standard basis vectors and several independent Gaussian vectors. The standard basis vectors were used to lower bound the trace $Tr(A^TA)$ where $A$ is the full embedding matrix (including any scaling factors), whereas the Gaussian vectors were used to upper bound the squared Frobenius norm $\|A^TA\|_F^2$. Since $Tr(A^TA)$ is the sum of the eigenvalues of $A^TA$ and $\|A^TA\|_F^2$ is the sum of squared eigenvalues, one cannot have a large $Tr(A^TA)$ and a small $\|A^TA\|_F^2$ without having many non-zero eigenvalues. Their lower bound on $m$ follows by observing that the number of non-zero eigenvalues equals the rank of $A$, and the rank cannot exceed $m$. We remark that the idea of using Gaussian vectors as a hard instance was also seen in \cite{DBLP:conf/approx/KaneMN11}. 

Compared to the proof above, we need to show something stronger. More precisely, the previous work merely showed that there are $\Omega(\eps^{-2} \lg n)$ non-zero eigenvalues. We need to show that there are $\Omega(\eps^{-2} \lg n)$ eigenvalues that are all at least $ds/(3m)$ large. This requires a more refined analysis and the introduction of the scaling parameter $s^{-1/2}$ in the embedding $s^{-1/2}Ax$ as in the definition of a \FastJL. 

The hard instance in our lower bound is also the standard basis vectors $e_1,\ldots,e_d$ in $\mathbb{R}^d$ together with a Gaussian distributed vector $g\in \mathbb{R}^d$. By Markov's inequality, we get that the following two events hold simultaneous with constant probability over the random choice of $A$: The number of basis vectors whose norm is preserved, i.e. $|\{i:\norm{Ae_i/\sqrt{s}} \in (1\pm\varepsilon)\}|$, is $\Omega(d)$, and secondly, the probability that the random Gaussian vector has its norm preserved satisfies $\ppg [\norm{Ag/\sqrt{s}}/\in(1\pm\eps)\norm{g} ]\geq 1-\Theta(\delta)$. Thus if we now consider an outcome $B$ of $A$ which satisfies these two relations, we get by $|\{i:\norm{Be_i/\sqrt{s}} \in (1\pm\varepsilon)\}|= \Omega(d)$ that the trace of $B^TB$, which is equal to the sum of the eigenvalues $B^TB$, is $\Omega(ds)$. Now by $\norm{Bg/\sqrt{s}}$ being in $(1\pm\eps)\norm{g}$ and $\norm{g}$ being in $(1\pm\eps)d$, both with probability least $1-\delta^{\Theta(1)}$ over $g$, we also get with probability at least $1-\delta^{\Theta(1)}$ over $g$ that $\norm{Bg}\in (1\pm\Theta(\eps))ds$. 

Now using the lower bound $\sum \lambda(B^TB)_i = \Omega(ds)$ and the fact that $B^TB$ has at most $m$ non-zero eigenvalues, we get that the sum of the eigenvalues larger than $ds/(3m)$ is at least $\Omega(ds)-m(ds/(3m)) =\Omega(ds)$ (provided that we can prove a large enough constant in the $\Omega(ds)$ notation). However, we also need to prove that there are not just a few such eigenvalues that are huge and account for most of the sum. For this, let $l$ denote the number of eigenvalues that are greater than or equal to $ds/(3m)$.

To prove a lower bound on $l$, we first use anti-concentration inequalities to relate the distribution of $\norm{Bg}$ to $Tr(B^TB)$, obtaining an upper bound on $\|B^TB\|_F^2 = \sum \lambda(B^TB)_i^2\leq O((ds)^2/m)$ (like in previous work). Using the upper bound on $\sum \lambda(B^TB)_i^2$ and Cauchy-Schwartz, we then conclude that the sum of the eigenvalues larger than $ds/(3m)$ is at most $\Theta(ds\sqrt{l/m})$ - hence combining the lower and upper bound on the sum of the eigenvalues larger than $ds/(3m)$, we get that $\Theta(ds\sqrt{l/m}) = \Omega(ds)$, so we conclude that $l=\Omega(m)$ as wanted. We remark that while this last part of our proof carries some resemblance to that in \cite{mlowerboundGreenNelson}, we believe that the whole reduction above, reducing the problem to arguing that the embedding matrix must have many large eigenvalues, is highly novel in its own right.

\paragraph{Preliminaries.}
To prove \Cref{lemma2}, we need the following two concentration bounds for normal distributed random variables.

\begin{lemma}\label{reformulation restatement 1} \cite{zhang2020nonasymptotic}
Let $g_1,\ldots,g_d$ be independent $N(0,1)$ random variables and $u_1,\ldots,u_d$ be non-negative numbers, then for constants $c_1\leq1$ and $C_1\geq 1$ we have that
$$
\begin{gathered}
c_1 \exp \left(-C_1 x^{2} /\|u\|_{2}^{2}\right) \leq \mathbb{P} \left [\sum_{i=1}^d u_i(g_i^2-1) \geq x\right ] , \quad \forall 0 \leq x \\
c_{1} \exp \left(-C_{1} x^{2} /\|u\|_{2}^{2}\right) \leq \mathbb{P}\left [\sum_{i=1}^du_i(g_i^2-1)\leq-x\right ] , \quad \forall 0 \leq x \leq c_1 \|u\|_{2}^{2}/\|u\|_{\infty}.
\end{gathered}
$$
\end{lemma}

\begin{lemma}\label{chisquarecons}(Example 2.11 \cite{wainwright_2019})
Let $g_1,\ldots,g_d$ be independent $N(0,1)$ random variables then 
$$
\mathbb{P}\left[\left| \sum_{k=1}^{d} g_i^{2}-d\right| \geq \alpha d\right] \leq 2 e^{-d \alpha^{2} / 8}, \quad \text { for all } \alpha \in(0,1).
$$
\end{lemma}

\paragraph{Proof of \Cref{lemma2}.}
We are now ready to give the proof of \Cref{lemma2}.
\begin{proof}
Let $A\in\mathbb{R}^{m\times d}$ be a \FastJL\ for $\varepsilon\leq1/4$ and $\delta\leq C$ (where $C$ is a constant to be fixed later), $t \varepsilon^{-2}\log(1/\delta)= m$ and $d\geq m$.

Let $e_1,\ldots,e_d$ be the standard basis vectors in $\mathbb{R}^d$. Let further $\ppg$ denote the measure of a standard Gaussian random vector $g\in\mathbb{R}^d$ independent of $A$. We now claim the existence of a set of matrices $S$ such that $A\in S$ holds with probability at least $1/2$ and for $B\in S$, we have that \begin{gather}\label{claim1}|\{i:|\norm{Be_i/\sqrt{s}}-\norm{e_i}|>\varepsilon\norm{e_i}\}|< 4\delta d
\end{gather}
and 
\begin{gather}\label{claim2}\ppg\left[|\norm{Bg/\sqrt{s}}-\norm{g}|>\varepsilon\norm{g}\right]< 4\delta.
\end{gather} 
To show this, define for each $i\in[d]$ the event $E_i=\{|\norm{Ae_i/\sqrt{s}}-\norm{e_i}|>\varepsilon\norm{e_i}\}$ and set $X_i$ equal to  $\mathbf{1}_{E_i}$, such that $\sum_{i=1}^d X_i=|\{i:|\norm{Be_i/\sqrt{s}}-\norm{e_i}|>\varepsilon\norm{e_i}\}|$. By the \FastJL\ assumption of $A$, we have that
$$\E_A\left[\sum_{i=1}^d X_i\right]\leq \delta d$$
so by Markov's inequality we get that
$$
\ppa\left[\sum_{i=1}^d X_i \geq 4\delta d\right]\leq 1/4
$$
similarly by the \FastJL\ assumption we have that 
$$\E_A\left[\ppg\left[|\norm{Ag/\sqrt{s}}-\norm{g}|>\varepsilon\norm{g}\right]\right]< \delta$$
so by applying Markov's inequality again, we get that
$$\ppa \left[\ppg\left[|\norm{Ag/\sqrt{s}}-\norm{g}|>\varepsilon\norm{g}\right]\geq 4\delta \right]\leq 1/4.$$
Now using a union bound gives that \cref{claim1} and \cref{claim2} hold simultaenously with probability at least $1/2$ as claimed.

If we can show that for $B\in S$, it holds that $\lambda(B^TB)_{\left \lceil cm/t\right \rceil} >ds/(3m)$, then we are done since the probability of $A$ being in $S$ is at least $1/2$. So let $B\in S$. We now notice that by \cref{claim1} there exist $(1-4\delta)d$ indices in $i\in[d]$ such that $(B^TB)_{i,i}\in(1\pm\varepsilon)s$. If we now let $\lambda_i(B^TB)$ denote the $i$'th largest eigenvalue of $B^TB$, we get the following lower bound on the sum of eigenvalues of $B^TB$ (assuming $\varepsilon\leq 1/4$ and $\delta\leq C\leq1/36$):
\begin{gather}\label{epsT}
\sum_{i=1}^m\lambda_i(B^TB)=Tr(B^TB)\geq(1-\eps)(1-4\delta) ds\geq 2ds/3.
\end{gather}
Now by Cauchy-Schwartz, we also have that
\begin{gather}\label{uppereigensum}
\sum_{i=1}^m\lambda_i(B^TB)\leq\sqrt{m\sum_{i=1}^m\lambda_i(B^TB)^2}\leq\sqrt{m\sum_{i=1}^m\lambda_i(B^TB)^2}\sqrt{\sum_{i=1}^m\lambda_i(B^TB)^2}/\lambda_1= \sqrt{m}\sum_{i=1}^m\lambda_i(B^TB)^2/\lambda_1
\end{gather}
Combining \cref{epsT} and \cref{uppereigensum}, we get that
\begin{gather}\label{lowerboundc_1}
(\sum_{i=1}^m \lambda_i(B^TB)^2)/\lambda_1(B^TB)\geq2ds/3\sqrt{m}\geq ds/4\sqrt{m}.
\end{gather}
Now since $B$ was in $S$, we have by \cref{claim2} that $\norm{Bg}\in(1\pm\varepsilon)\norm{g}$ with probability at least $1-4\delta$ over $g$.
At the same time, we have by \Cref{chisquarecons} that for $0<\alpha<1$, it holds that $\norm{g}\in (1\pm \alpha)d$ with probability at least $1-2\exp{(-d\alpha^2/8)}$. Now choosing $\alpha=\varepsilon$, we get that $2\exp(-d\varepsilon^2/8)\leq 2\delta^{1/8}$. By the assumption that $d\geq m\geq \eps^{-2}\log(1/\delta)$, we get that $\norm{g}\in(1\pm\varepsilon)d$ with probability at least $1-2\delta^{1/8}$ over $g$. Now combining this with $\norm{Bg}\in(1\pm\varepsilon)\norm{g}$ with probability at least $1-4\delta$ over $g$, we get by a union bound that
\begin{gather}\label{agbound}
\norm{Bg}\in(1\pm\varepsilon)(1\pm\varepsilon)ds=(1-2\varepsilon+\varepsilon^2,1+2\varepsilon+\varepsilon^2)ds
\end{gather}
with probability at least $1-6\delta^{1/8}$ over $g$.

Now using the eigenvalue decomposition of $B^TB$ into $U^TDU$, where $U$ is an orthogonal matrix and $D$ an diagonal matrix with the eigenvalues of $B^TB$ on its diagonal in decreasing order, and that a standard normal Gaussian vector is invariant in distribution under rotations, we obtain the following relation
\begin{gather}
\norm{Bg}-Tr(B^TB) = \nonumber\\
g^T B^TBg-Tr(B^TB) =\nonumber\\
g^TU^TDUg-Tr(B^TB) \stackrel{d}{=} \label{normaschi}\\
\tilde{g}^TD\tilde{g}-\sum_{i=1}^d\lambda_i(B^TB) = \nonumber\\
\sum_{i=1}^d\lambda_i(B^TB)(\tilde{g}_i^2-1)\nonumber.
\end{gather}
Our next step is to relate $\sum_i \lambda_i^2(B^TB)$ to $\delta$. Here we take two different approaches depending on $Tr(B^TB)$. $c_1$ and $C_1$ in the following are the constants of \Cref{chisquarecons}.

\paragraph{Case 1:} If $Tr(B^TB)\leq( 1-2\varepsilon+c_1/(4\sqrt{m}))ds$ then by \cref{agbound} (and the comment above the equation) we have with probability at least $1-6\delta^{1/8}$ over $g$ that
\begin{gather*}\norm{Bg}-Tr(B^TB)\geq ((1-2\varepsilon+\varepsilon^2)-( 1-2\varepsilon+c_1/(4\sqrt{m})))ds>- c_1ds/4\sqrt{m}.
\end{gather*}
implying that $6\delta^{1/8}\geq  \ppg\left[\norm{Bg}-Tr(B^TB)\leq-c_1ds/4\sqrt{m} \right]$. 

Now noticing that $c_1ds/4\sqrt{m} \leq c_1(\sum_{i=1}^m \lambda_i(B^TB)^2)/\lambda_1(B^TB)$ by \cref{lowerboundc_1},  we may invoke the second relation in \Cref{reformulation restatement 1} on \cref{normaschi} to get:
\begin{gather*}
 \ppg\left[\norm{Bg}-Tr(B^TB)\leq-c_1ds/4\sqrt{m} \right]\\
=\ppgt\left[\sum_{i=1}^d\lambda_i(B^TB)(\tilde{g}_i^2-1)\leq-c_1ds/4\sqrt{m} \right]\geq c_1 \exp\left(-C_1(c_1ds)^2/(16m\sum_{i=1}^d\lambda_i^2(B^TB))\right).
\end{gather*}
Yielding that $6\delta^{1/8}\geq c_1 \exp\left(-C_1(c_1ds)^2/(16m\sum_{i=1}^d\lambda_i^2(B^TB))\right)$.

\paragraph{Case 2:}
If $Tr(B^TB)\in [( 1-2\varepsilon+c_1/(4\sqrt{m}))ds,\infty)$ then by \cref{agbound} (and the comment below the equation) we have with probability at least $1-6\delta^{1/8}$ over $g$ that
\begin{gather*}
\norm{Bg}-Tr(B^TB)
\leq ((1+2\varepsilon+\varepsilon^2)-(1-2\varepsilon+c_1/(4\sqrt{m}))))ds
< 5\varepsilon ds.
\end{gather*}
implying that $6\delta^{1/8}\geq  \ppg\left[\norm{Bg}-Tr(B^TB)\geq 5\eps ds\right]$.

Now using the first relation in \Cref{reformulation restatement 1} combined with \cref{normaschi}, it follows that
\begin{gather*}
 \ppg\left[\norm{Bg}-Tr(B^TB)>5\varepsilon ds\right]\\
=\ppgt\left[\sum_{i=1}^d\lambda_i(B^TB)(\tilde{g}_i^2-1)>5\varepsilon ds\right]\geq c_1 \exp\left(-C_1(5\varepsilon ds)^2/(\sum_{i=1}^d\lambda_i^2(B^TB))\right).
\end{gather*}
Yielding that $6\delta^{1/8}\geq c_1 \exp\left(-C_1(5\varepsilon ds)^2/(\sum_{i=1}^d\lambda_i^2(B^TB))\right)$.

\paragraph{Conclusion.}
Now using that $m\geq\eps^{-2} \log(1/\delta)$ and $c_1\leq 1$ it follows that $c_1^2/16m\leq 5^2\eps^2$ which then implies that $C_1(c_1ds)^2/(16m\sum_{i=1}^d\lambda_i^2(B^TB))\leq C_1(5\varepsilon ds)^2/(\sum_{i=1}^d\lambda_i^2(B^TB)$. Combining this with the conclusion of the above two cases, we get that $6\delta^{1/8}\geq c_1 \exp\left(-C_1(5\varepsilon ds)^2/(\sum_{i=1}^d\lambda_i^2(B^TB))\right)$. With this relation, choosing the universal constant  $C= (c_1/6)^{16}$ (less than $1/36$ as used in \cref{epsT}), which implies that $c_1/(6\delta^{1/16})\geq 1$, and using that $m = t\eps^{-2}\log(1/\delta)$, we now get that
\begin{gather}
\log(6\delta^{1/8})\geq \log(c_1)-C_1(5\varepsilon ds)^2/(\sum_{i=1}^d\lambda_i^2(B^TB))\nonumber \\
\Rightarrow \sum_{i=1}^d\lambda_i^2(B^TB)\leq C_1(5\varepsilon ds)^2/(\log(c_1/(6\delta^{1/8})))\leq C_116(5\varepsilon ds)^2/\log(1/\delta)\leq 20^2C_1t(ds)^2/m\label{bound forbenuis}
\end{gather}
We now define the vector $w\in\mathbb{R}^d$ as
\[
[w]_i =\begin{cases} 

   1 & \text{if } \lambda_i(B^TB)\geq ds/(3m)\\
&\\  % blank row
0
    & \text{else } 
\end{cases}
\]
and let $l$ be equal to the number of non-zero entries of $w$. Let further $\lambda $ denote the vector in $\mathbb{R}^d$ with the eigenvalues of $B^TB$ in decreasing order. It then follows by Cauchy-Schwartz and \cref{bound forbenuis} that we have the following upper bound on the sum of the eigenvalues of $B^TB$ larger than $ds/(3m)$:
\begin{gather*}
\sum_{i: \lambda_i(B^TB)\geq ds/(3m)}\lambda_i(B^TB)=\left\langle \lambda,w\right\rangle\leq \normos{\lambda}\normos{w}=\sqrt{\sum_{i=1}^d\lambda_i^2(B^TB)l}\leq \sqrt{20^2C_1t(ds)^2l/m}.
\end{gather*}
At the same time, we get the following lower bound on the sum of the eigenvalues of $B^TB$ larger than $ds/(3m)$  by \cref{epsT} and the fact that $(B^TB)$ has rank at most $m$ and hence at most $m$ non-zero eigenvalues
\begin{gather*}
\sum_{i: \lambda_i(B^TB)\geq ds/(3m)}\lambda_i(B^TB)=\sum_{i=1}^d\lambda_i(B^TB)-\sum_{i: \lambda_i(B^TB)< ds/(3m)}\lambda_i(B^TB)\geq 2ds/3-ds/3=ds/3.
\end{gather*}
Hence combining the upper and lower bound we obtain that $ds/3\leq \sqrt{20^2C_1t(ds)^2 l/m}$, implying that $m/(60^2C_1t)\leq l$, which by setting $c$ in \Cref{lemma2} equal to $1/60^2C_1\leq1$ ($C_1\geq1$ by \Cref{reformulation restatement 1}) concludes the proof of \Cref{lemma2}.
\end{proof}

\bibliographystyle{abbrv}
\bibliography{bibliography}

\end{document}